\documentclass{article}
\usepackage{graphicx}
\usepackage{amsmath}
\usepackage{amssymb}
\begin{document}

\begin{center}
{\Large{Bidirectional Whitham Equations as Models of Waves on Shallow Water}}\\
{\large{John D.~Carter}}\\
carterj1@seattleu.edu\\
April 30, 2018

\end{center}

\section{Abstract}

Hammack \& Segur~\cite{HamSeg} conducted a series of surface water-wave experiments in which the evolution of long waves of  depression was measured and studied.  This present work compares time series from these experiments with predictions from numerical simulations of the KdV, Serre, and five unidirectional and bidirectional Whitham-type equations.  These comparisons show that the most accurate predictions come from models that contain accurate reproductions of the Euler phase velocity, sufficient nonlinearity, and surface tension effects.  The main goal of this paper is to determine how accurately the bidirectional Whitham equations can model data from real-world experiments of waves on shallow water.  Most interestingly, the unidirectional Whitham equation including surface tension provides the most accurate predictions for these experiments.  If the initial horizontal velocities are assumed to be zero (the velocities were not measured in the experiments), the three bidirectional Whitham systems examined herein provide approximations that are significantly more accurate than the KdV and Serre equations.  However, they are not as accurate as predictions obtained from the unidirectional Whitham equation.

\section{Introduction}
\label{intro}

Hammack \& Segur~\cite{HamSeg} performed a series of
tightly-controlled laboratory water-wave experiments in a long, narrow
tank with relatively shallow (10~cm) undisturbed water and a wave
maker at one end.  The wave maker was a rectangular, vertically-moving
piston located on the bottom of the tank, adjacent to a rigid wall at
the upstream end of the tank.  Experiments were initialized by rapidly
moving the piston downward a prescribed amount that varied between
experiments.  This downward motion lead to the creation of initially
rectangular waves wholly below the still water level, occupying the
entire width and 61 cm of the upstream end of the tank.  The evolution
of the wave train downstream from the wave maker was investigated.
Time series were collected at five gauges located 61~cm ($x=0$, the
downstream edge of the piston), 561~cm ($x=500$), 1,061~cm ($x=1000$),
1,561~cm ($x=1500$), and 2,061~cm ($x=2000$) downstream.  The tank was
long enough that waves reflecting from the far end of the tank did not
impact the time series collected.  Among other things, Hammack \&
Segur showed that many analytic and asymptotic results obtained from
the KdV equation compared favorably with measurements from the
experiments.

In this paper, we focus on the two experiments presented in Figures 2
and 3 of~\cite{HamSeg}, which we refer to as experiment \#2 and
experiment \#3 respectively.  The experiments were identical except
for the magnitude of the piston displacement and hence initial wave
amplitude. In experiment \#2, the piston moved downward 1~cm,
producing a downstream propagating wave with an initial amplitude of
0.5~cm.  In experiment \#3, the piston stroke and initial amplitude
were 3~cm and 1.5~cm respectively.  The time series from both
experiments show leading triangular waves of depression followed by
series of trailing wave groups.

The main goal of this paper is to compare and evaluate a number of unidirectional and bidirectional Whitham-type equations by comparing their predictions with the experimental time series.  In doing this, we demonstrate that in order to most accurately model these experimental measurements, a model must include (i) an accurate reproduction of the Euler phase velocity, (ii) sufficient nonlinearity, and (iii) surface tension effects.  To our knowledge, these are the first comparisons between the recently derived bidirectional Whitham equations and data from physical experiments.

This paper is organized as follows.  The model equations and their
properties are presented in Section \ref{eqns}.  Comparisons between
the experimental time series and data from numerical simulations of
these equations are included in Section \ref{numerics}.  A summary is
contained in Section \ref{summary}

\section{Model Equations}
\label{eqns}

The equations that describe the irrotational motion of an inviscid,
incompressible, homogeneous fluid with a free surface are known as the
Euler equations~\cite{johnson}.  As the experiments of interest here
were conducted in a long, narrow tank, we use two-dimensional models
(i.e.~models with one horizontal and one vertical dimension).  The
linear phase velocity for the Euler equations is given by
\begin{equation}
c_{_E}=\pm\sqrt{\frac{(g+\tau k^2)\tanh(kh_0)}{k}},
\label{EulerDisp}
\end{equation}
where $g$ represents the acceleration due to gravity, $\tau$ represents the coefficient of surface tension, $h_0$ represents the mean depth of the fluid, and $k$ represents the wave number of the linear wave.  In all of our calculations we used $g=981$~cm/sec$^2$ and $\tau=72.86$~cm$^3$/sec$^2$, the surface tension of pure water at 20 degrees Celsius~\cite{SurfaceTension}.   The plus or minus sign in equation (\ref{EulerDisp})
establishes that the Euler equations are bidirectional (waves of each
wave number can travel toward both $x=-\infty$ and  $x=\infty$
as $t$ increases) as opposed to unidirectional (waves of each wave
number travel toward only $x=-\infty$ or $x=\infty$ as $t$
increases).  Since the Euler equations are difficult to work with, it
is common to introduce the dimensionless parameters
\begin{equation}
\delta=\frac{h_0}{\lambda_0},\hspace*{2cm}\epsilon=\frac{a_0}{h_0},
\end{equation}
in order to derive asymptotic models that are less complicated.  Here
$\lambda_0$ is a typical wavelength and $a_0$ is a typical wave amplitude.
The parameter $\epsilon$ is a measure of
nonlinearity and the parameter $\delta$ is a measure of wavelength or
shallowness.

\subsection{The KdV Equation}
\begin{figure}
\centering
\includegraphics[width=8cm]{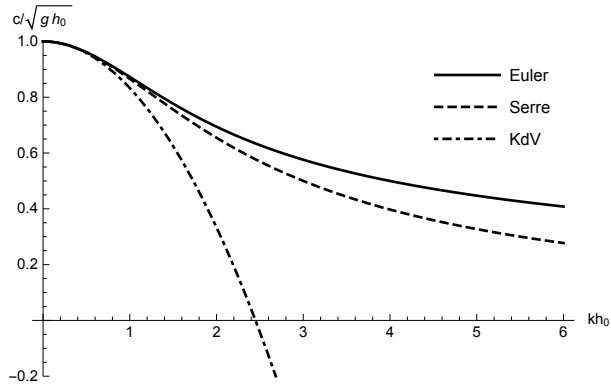}
\caption{Plots of scaled linear phase velocity versus scaled wave number
  for the Euler, KdV, and Serre equations without surface tension.}
\label{dispfig}
\end{figure}

The Korteweg-de Vries (KdV) equation can be derived from the Euler
equations by assuming that $\delta^2\sim\epsilon\ll 1$ and truncating
at $\mathcal{O}(\epsilon^3)$.  In other words, the waves are assumed
to have small amplitude and large wavelength.  The KdV equation has
been well studied mathematically (e.g.~\cite{miles,Whithambook,AS,LannesBook}) and
experimentally (e.g.~\cite{russell,zabusky,h,hs}).  In dimensional
form, the KdV equation with surface tension~\cite{LannesBook} is given by
\begin{equation}
\eta_t+\sqrt{gh_0}~\eta_x+\frac{3}{2h_0}\sqrt{gh_0}~\eta\eta_x+\sqrt{gh_0}\Big{(}\frac{h_0^2}{6}-\frac{\tau}{2g}\Big{)}\eta_{xxx}=0,
\label{KdV}
\end{equation}
where $\eta=\eta(x,t)$ represents the displacement of the free surface
from its undisturbed level. The linear phase velocity for KdV with surface tension is
\begin{equation}
c_{_{K}}=\sqrt{gh_0}\Big{(}1-k^2\big{(}\frac{h_0^2}{6}-\frac{\tau}{2g}\big{)}\Big{)}.
\end{equation}
Figure \ref{dispfig} contains a plot of the phase velocity for KdV without surface tension (i.e.~$\tau=0$).  The
plot establishes that KdV only accurately approximates Euler's phase
velocity near $kh_0=0$ (i.e.~in the long-wave limit).  This establishes that KdV is a weakly dispersive model.  Additionally, for a given $k$, there is a unique $c_{_{K}}$, so KdV is a unidirectional model (even though $k$ such that $|k|<\sqrt{\frac{6g}{gh_0^2-3\tau}}$ travel toward $x=\infty$ and $k$ such that $|k|>\sqrt{\frac{6g}{gh_0^2-3\tau}}$ travel toward
$x=-\infty$).

\subsection{The Serre Equations}
\label{Serresection}

The first strongly nonlinear, weakly dispersive set of Boussinesq-type
equations was derived by Serre~\cite{serre,serre2}.  Several years
later, Su \& Gardner~\cite{SG} and Green \& Naghdi~\cite{GN}
re-derived these equations using different methods.  Many have presented rigorous, perturbation theory derivations of the Serre equations, see for example Johnson~\cite{johnsonJFM} and Lannes~\cite{LannesBook}.  In the literature, the equations are
referred to as both the Serre and the Green \& Naghdi
equations.  They are obtained by depth-averaging the Euler system and
truncating the resulting set of equations at $\mathcal{O}(\delta^4)$
without making any assumptions on $\epsilon$.  This ``full
nonlinearity'' makes the Serre equations ideal for studying
large-amplitude or nearly breaking waves.  The
dimensional Serre equations are given by
\begin{subequations}
\begin{equation}
h_t+(h\bar{u})_x=0,
\end{equation}
\begin{equation}
\bar{u}_t+\bar{u}\bar{u}_x+gh_x-\tau h_{xxx}-\frac{1}{3h}\Big{(}h^3\big{(}\bar{u}_{xt}+\bar{u}\bar{u}_{xx}-(\bar{u}_x)^2\big{)}\Big{)}_x=0,
\end{equation}
\label{Serre}
\end{subequations}
where $\bar{u}=\bar{u}(x,t)$ represents the depth-averaged horizontal
velocity of the fluid and $h=h(x,t)$ represents the local fluid
depth.  Fluid depth is related to surface displacement via
\begin{equation}
h(x,t)=h_0+\eta(x,t).
\label{heta}
\end{equation}
The Serre equations are bidirectional, dispersive, and have the
following linearized phase velocity
\begin{equation}
c_{_{S}}=\pm\sqrt{\frac{3h_0(g+\tau k^2)}{3+(kh_0)^2}}.
\end{equation}
Figure \ref{dispfig} shows that the Serre equations without surface tension do a better job of approximating the Euler phase velocity than KdV, but they only accurately reproduce the Euler phase velocity for $|k|h_0 \lesssim 1.4$.

\subsection{The Whitham Equation}
\label{Whithamsection}
Neither KdV nor Serre accurately reproduces linearized phase velocity
of the Euler equations for a wide range of $kh_0$ values.  In
order to address this issue, Whitham~\cite{Whitham,Whithambook}
proposed a generalization of KdV that is now known as the Whitham
equation for waves on shallow water.  In dimensional form, it is given
by
\begin{equation}
\eta_t+\sqrt{gh_0}~\mathcal{K}\eta_x+\frac{3}{2h_0}\sqrt{gh_0}~\eta\eta_x=0,
\label{Whitham}
\end{equation}
where $\mathcal{K}$ is a (dimensionless) Fourier multiplier defined by
the symbol
\begin{equation}
\widehat{\mathcal{K}f}(k)=\sqrt{\frac{(1+\frac{\tau}{g}k^2)\tanh(kh_0)}{kh_0}}~\hat{f}(k).
\label{Kappa}
\end{equation}
The linear phase velocity for the Whitham equation is
\begin{equation}
c_{_W}=\sqrt{\frac{(g+\tau k^2)\tanh(kh_0)}{k}}.
\end{equation}
The Whitham equation provides a unidirectional reproduction of the Euler phase velocity.  We show in the next section that the more accurate the reproduction of the Euler phase velocity causes the Whitham equation to more accurately predict the evolution of waves than the KdV equation (which has the same nonlinearity, but less accurate dispersion).  In the last decade, the Whitham equation has received significant attention in the mathematics community.  We outline the results from some of this work in the remainder of this paragraph.  Ehrnstr\"om \& Kalisch~\cite{EK} proved the
existence of and computed periodic traveling-wave solutions to the
Whitham equation.  Sanford {\emph{et al.}}~\cite{WhithamStability} and
Johnson \& Hur~\cite{MatVera} established that large-amplitude,
periodic traveling-wave solutions to the Whitham equation are unstable
while small-amplitude, periodic traveling-wave solutions are stable if
the wavelength is long enough.   Moldabayev {\emph{et
    al.}}~\cite{Moldabayev} presented a scaling regime in which the
Whitham equation can be derived and compared its dynamics with those
from other models including the Euler equations.
Hur~\cite{WhithamBreaking} proved that solutions to the Whitham
equation will break provided that the initial condition is
sufficiently asymmetric.  Deconinck \& Trichtchenko~\cite{BernardOlga}
proved that the unidirectional nature of the Whitham equation causes
it to miss some of the instabilities of the Euler equations.  Dinvay {\emph{et al.}}~\cite{Dinvay} extend the word of Moldabayev {\emph{et al.}}~\cite{Moldabayev} to include surface tension and show that the Whitham equation including surface tension gives a more accurate reproduction of the free-surface problem than the KdV and Kawahara equations.
To our
knowledge, only Trillo {\emph{et al.}}~\cite{Trillo} has compared Whitham
predictions with measurements from laboratory experiments.  They
showed that the Whitham equation provides an accurate model for the
evolution of initial waves of depression, especially when nonlinearity
plays a significant role.

In addition to equation (\ref{Whitham}), Whitham also proposed the following equation
\begin{equation}
\eta_t+\sqrt{gh_0}~\mathcal{K}\eta_x+3\Big{(}\sqrt{g(h_0+\eta)}-\sqrt{gh_0}\Big{)}\eta_x=0.
\label{SqrtWhitham}
\end{equation}
We refer to this equation as the Sqrt Whitham equation.  It provides the same unidirectional reproduction of the Euler phase velocity as does the Whitham equation.  The first term in the $\eta=0$ Taylor series expansion of the terms in parentheses gives the nonlinear term in the Whitham equation. This means that the Sqrt Whitham equation has ``more" nonlinearity than the Whitham equation.


\subsection{Bidirectional Whitham Systems}
\label{BidWhitham}

Boussinesq-type models are obtained from the Euler equations in the
long-wave, small-amplitude limit.  There is a large number of
Boussinesq-type systems (e.g.~\cite{Bous,BCS,LannesBook}).  For
example, the KdV equation is a unidirectional version of the Boussinesq
system~\cite{johnson}.  There are multiple ``Whithamized''
Boussinesq models, that is Boussinesq-type systems that have been
modified so that their phase velocities match the bidirectional phase
velocity of the Euler equations.  To our knowledge, this work contains the first comparisons between predictions from bidirectional Whitham systems and data from physical experiments.

Aceves-S\'anchez {\emph{et al.}}~\cite{AcevesSanchez} constructed the
following Hamiltonian, bidirectional Whitham system
\begin{subequations}
\begin{equation}
\eta_t+h_0\mathcal{K}^2u_x+(\eta u)_x=0,
\end{equation}
\begin{equation}
u_t+g\eta_x+uu_x=0,
\end{equation}
\label{ASMP}
\end{subequations}
where $\mathcal{K}$ is defined in equation (\ref{Kappa}).  We refer to
this system as the ASMP system.  It has the following conserved
quantities
\begin{subequations}
\begin{equation}
\mathcal{Q}_1=\int_0^L\eta~dx,
\end{equation}
\begin{equation}
\mathcal{Q}_2=\int_0^Lu~dx,
\end{equation}
\begin{equation}
\mathcal{Q}_3=\int_0^L\eta u~dx,
\end{equation}
\begin{equation}
\mathcal{Q}_4=\int_0^L\frac{1}{2}\Big{(}g\eta^2+h_0u\mathcal{K}^2u+\eta
u^2\Big{)}dx,
\label{Q4}
\end{equation}
\end{subequations}
where $L$ is the spatial period of the solution.  The fourth conserved
quantity, $\mathcal{Q}_4$, is the Hamiltonian of the system.  This
system is bidirectional and has a linear phase velocity that exactly
matches that of the Euler equations (i.e.~it is ``fully
dispersive'').  Ehrnstr\"om {\emph{et al.}}~\cite{EhrnstromPeiWang} prove that the ASMP system is well-posed if the surface displacement is non-vanishing and suggest that it is ill-posed in general.  Claassen \& Johnson~\cite{Kyle} present numerical results that numerically corroborate these well/ill-posedness statements.
Moldabayev {\emph{et al.}}~\cite{Moldabayev} presented a scaling
regime in which the ASMP system can be derived from the Euler
equations.  Additionally, they extended their procedure and derived a
fully dispersive Hamiltonian system with higher-order nonlinearity.
This system is given by
\begin{subequations}
\begin{equation}
\eta_t+h_0\mathcal{K}^2u_x+(\eta u)_x-\delta^2h_0^2(\eta u_x)_{xx}=0,
\end{equation}
\begin{equation}
u_t+g\eta_x+uu_x+\delta^2h_0^2u_xu_{xx}=0.
\end{equation}
\label{MKD}
\end{subequations}
We refer to this system as the MKD system.  In our numerical simulations, we used the wavelength of the initial surface displacement to define
$\delta$.  It is
bidirectional, fully dispersive and conserves $\mathcal{Q}_1$,
$\mathcal{Q}_2$, $\mathcal{Q}_3$, and its Hamiltonian,
\begin{equation}
\mathcal{Q}_5=\int_0^L\frac{1}{2}\Big{(}g\eta^2+h_0u\mathcal{K}^2u+\eta
u^2+\delta^2h_0^2\big{(}\eta_xuu_x+\eta uu_{xx} \big{)} \Big{)}dx.
\end{equation}

Another bidirectional Whitham system was proposed by Hur \& Pandey~\cite{HurPandey} and Hur \& Tao~\cite{HurTao}.  In dimensional form, this system is given
by
\begin{subequations}
\begin{equation}
\eta_t+h_0u_x+(\eta u)_x=0,
\end{equation}
\begin{equation}
u_t+g\mathcal{K}^2\eta_x+uu_x=0.
\end{equation}
\label{HP}
\end{subequations}
We refer to this system as the HP system.  It is a fully dispersive bidirectional model.  Hur \& Pandey established the system is well-posed for short times and that periodic traveling-wave solutions are spectrally unstable with respect to long-wave perturbations.  The HP system conserves $\mathcal{Q}_1$, $\mathcal{Q}_2$, $\mathcal{Q}_3$, and its Hamiltonian,
\begin{equation}
\mathcal{Q}_6=\int_0^L\frac{1}{2}\Big{(}h_0u^2+g\eta\mathcal{K}^2\eta+\eta u^2\Big{)}dx,
\end{equation}
but not $\mathcal{Q}_4$ or $\mathcal{Q}_5$.  The HP system is asymptotically equivalent to the ASMP system in the KdV regime.

\section{Numerics and Comparisons}
\label{numerics}

\subsection{Initial conditions}

\begin{figure}
\centering
\includegraphics[width=6cm]{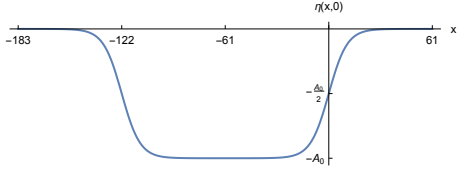}
\caption{A plot of the non-zero portion of the initial surface
  displacement given in equation (\ref{ICs}).}
\label{ICplot}
\end{figure}

For simplicity, we ignored the motion of the piston, assumed that the
bottom of the tank is horizontal, and assumed that the initial wave
profile is a nearly rectangular wave of depression.  Additionally,
we assumed periodic boundary conditions on an interval large enough
that waves did not wrap around and erroneously influence the
numerical gauges.  The initial conditions used for the numerical
simulations of the KdV, Whitham, and Sqrt Whitham equations were
\begin{equation}
\eta(x,0)=\left\{
     \begin{array}{lr}
       0 & -7869\le x<-183,\\
       -\frac{1}{2}A_0+\frac{1}{2}A_0\mbox{sn}(0.0925434x,0.9999^2) & -183\le x\le 61,\\
       0 & 61<x\le 7747,
     \end{array}
   \right.
\label{ICs}
\end{equation}
where $A_0$ is the amplitude of the piston motion ($A_0=0.5$ in
experiment \#2 and $A_0=1.5$ in experiment \#3) and
$\mbox{sn}(\cdot,m)$ is a Jacobi elliptic function with elliptic modulus $m$~\cite{BF}.  A plot of the nonzero portion of this initial condition is included in Figure \ref{ICplot}.  These initial conditions were chosen because they accurately approximate the even extension of the (dimensional) experimental initial conditions.  The position $x=0$ in the numerical tank corresponds to the rightmost/downstream edge of the experimental piston and the location of the first gauge.  The parameters in the initial conditions were chosen so that numerical gauges could be placed at $x=0$ and close to $x=500, 1000, 1500, 2000$ (the locations of the experimental gauges).

Due to the relation between $h$ and $\eta$ (see equation (\ref{heta})), the initial conditions used for simulations of the
Serre equations, (\ref{Serre}), were
\begin{subequations}
\begin{equation}
h(x,0)=10+2\eta(x,0),
\end{equation}
\begin{equation}
\bar{u}(x,0)=u(x,0)=0,
\end{equation}
\label{ICs2}
\end{subequations}
where $\eta(x,0)$ is defined in equation (\ref{ICs}).  Note that the mean fluid depth for both experiments was 10~cm, so $h_0=10$.  The amplitude factor of $2$ in the surface displacement is necessary because the Serre equations are bidirectional while the experiments, the KdV, Whitham, and Sqrt Whitham equations are unidirectional.    Unfortunately the horizontal velocities were not measured in the experiments.  Therefore, it is unclear how to properly choose the initial horizontal velocities, especially since the fluid motion was initiated by a vertically moving piston. To avoid confusion, we set the initial horizontal fluid velocities to zero.  The impact of this choice (and others) is discussed in Section \ref{BidWhNumerics}.

Since the ASMP, (\ref{ASMP}), the MKD, (\ref{MKD}), and the HP, (\ref{HP}), equations are bidrectional, the initial surface displacement for these equations needs to be twice the initial surface displacement given in equation (\ref{ICs}).  The initial horizontal velocities for these equations were chosen to be zero just as was done with the Serre system.

The Serre equations were solved using the iterative pseudospectral
method presented by Dutykh {\emph{et al.}}~\cite{DCMM}.  All other
models were solved using sixth-order split-step~\cite{yoshida}
pseudospectral methods where the linear and nonlinear  parts were
solved independently.  This allowed the linear parts of the PDEs
(including the linear parts of the fully dispersive models) to be
solved exactly using fast Fourier transforms.  A three-eighths rule was
required to solve the MKD system for experiment \#3 and the Serre system for both experiments, but was not used
in any of the other simulations.  The conserved quantities were used as checks on the accuracy of the numerical methods.

\subsection{Results}

In order to quantitatively compare the predictions from the various
models, we used the norm
\begin{equation}
\mathcal{E}=\frac{1}{5}\sum_{k=1}^{5}\frac{\sum_{j=1}^{N_k}\big{|}\mbox{expt}(k,j)-\mbox{num}(k,j)\big{|}^2}{\sum_{j=1}^{N_k}\big{|}\mbox{expt}(k,j)\big{|}^2},
\label{error}
\end{equation}
where $N_k$ is the number of experimental temporal data points at gauge $k$ (there were five gauges), and $\mbox{expt}(k,j)$ and $\mbox{num}(k,j)$ are the $j^{th}$ experimental and numerical data values for the $k^{th}$ gauge respectively.  We did not use a simple percent error formula because there are times at which the experimental value for a gauge are zero/close to zero (which would have led to division by zero/small numbers).  As the experimental data was redigitized, we aligned the experimental time series and numerical predictions by aligning the first wave of elevation at the first gauge.  Table \ref{ErrorTable} contains $\mathcal{E}$ values for the model equations and both experiments.

\begin{table}
\begin{center}
\begin{tabular}{|ccc|}
\hline
PDE & $\mathcal{E}$ expt. 2 & $\mathcal{E}$ expt. 3 \\
\hline
Linear Theory & 0.6294 & ---\\
KdV with $\tau=0$ & 0.1238 & 1.6828\\
KdV  & 0.1226 & 1.6813\\
Serre with $\tau=0$ & 0.1048 & 1.4209\\
Serre & 0.1048 & 1.4133\\
Whitham with $\tau=0$ & 0.0734 & 0.9363\\
Whitham & 0.0717 & 0.9212\\
Sqrt Whitham & 0.0716 & 0.9740\\
ASMP with $\tau=0$ & 0.0788 & 1.1090\\
ASMP & 0.0777 & 1.1119\\
HP with $\tau=0$ & 0.0798 & 1.2316\\
HP & 0.0786 & 1.2056\\
MKD & 0.0735 & 0.9799\\
\hline
\end{tabular}
\end{center}
\caption{Error values, see equation (\ref{error}), for the model equations and both experiments.}
\label{ErrorTable}
\end{table}

\subsubsection{The KdV Equation}

\begin{figure}
\centering
\includegraphics[width=16cm]{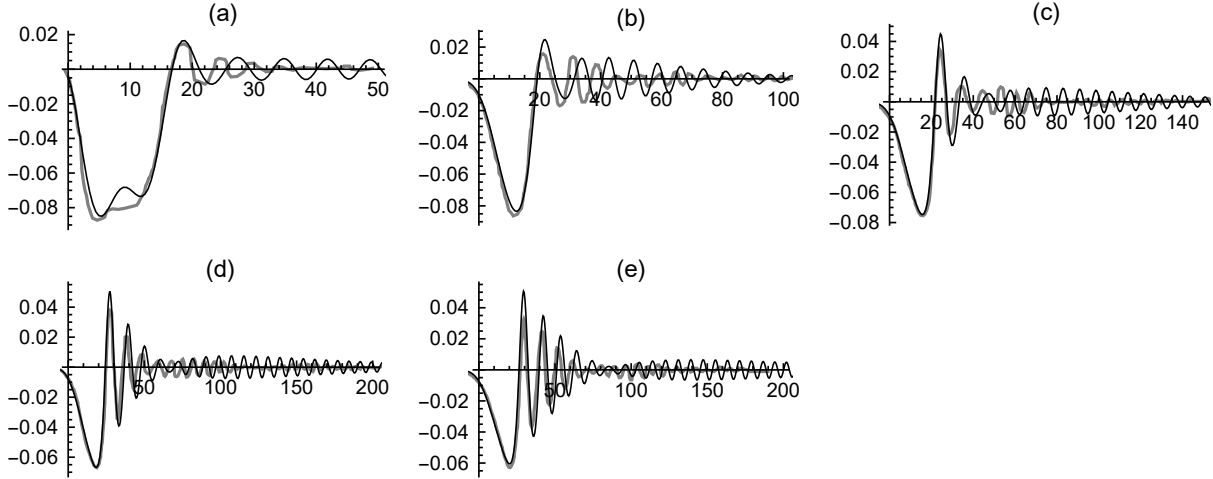}
\caption{Plots comparing results from a numerical solution of KdV including surface tension
  (thin, black curves) with experimental data (thick, gray curves)
  corresponding to experiment \#2.  The vertical axes are scaled
  surface displacement, $\frac{3}{2}\frac{\eta}{h_0}$, and the
  horizontal axes are scaled time,
  $\sqrt{\frac{g}{h_0}}~t-\frac{x}{h_0}$, with
  $\frac{x}{h_0}=0,~50,~100,~150,~200$ for (a), (b), (c), (d), and (e)
  respectively.}
\label{KdV2}
\end{figure}

Figure \ref{KdV2} contains plots that compare results from numerical simulations of KdV including surface tension with the experimental data for experiment \#2 at each of the five gauges.  Figure \ref{KdV2}(a) shows that, at the gauge closest to the wave maker, KdV does a reasonable job modeling the amplitudes of the initial wave of depression and first wave of elevation, but overestimates the amplitudes of the remaining trailing waves.  Figures \ref{KdV2}(b)-(e) include the results from the four downstream gauges.  They show that KdV accurately predicts the amplitude of the initial wave of depression, but overestimates the amplitudes of the trailing waves of elevation.  This amplitude overestimation may be due to experimental dissipative effects that KdV (a nondissipative model) does not account for.

At each gauge, KdV accurately models the phases of the first two waves
(the initial wave of depression and the initial wave of elevation),
but does not accurately model the phases of any of the trailing waves.
As time increases at each gauge, this phase error increases.  This is
consistent with the fact that KdV accurately models the fastest waves and underestimates the speeds of the slower waves.  Finally, KdV completely misses the group nature
of the trailing dispersive waves.

Table \ref{ErrorTable} shows that the KdV equation with surface tension does slightly better in modeling experiment \#2 than KdV without surface tension.  More importantly, the table shows that KdV did not perform well in comparison with the other models according to the error norm given in equation (\ref{error}).  The only model that KdV outperformed is the ``linear theory'' model.  The linear theory model is simply
the (bidirectional) linearized Euler system.  It does a poor job
modeling the data from experiment \#2 and does not even qualitatively
predict the wave evolution in experiment \#3 (and therefore its result
is left out of Table \ref{ErrorTable}).  The poor performance of the
linear theory suggests that some form of nonlinearity is required in
order to accurately model data from these experiments.  The role of
nonlinearity is discussed further in Sections \ref{SerreNumerics} and
\ref{BidWhNumerics}.  The poor performance of KdV in modeling the phases
suggests that a model with more accurate dispersion needs to be used.
The role of dispersion is further discussed in each of the remaining
sections.

\begin{figure}
\centering
\includegraphics[width=16cm]{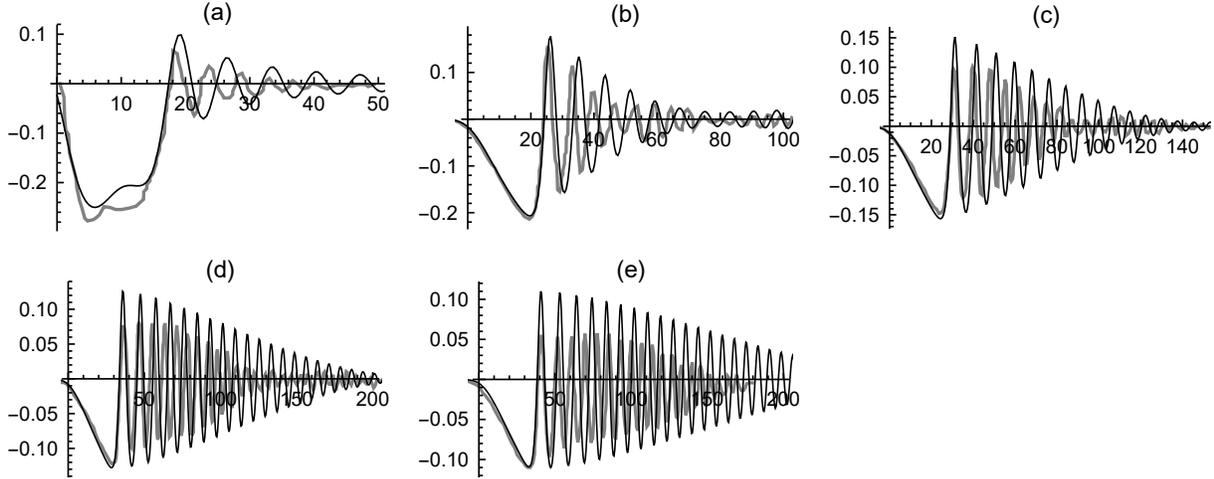}
\caption{Plots comparing results from a numerical solution of KdV including surface tension
  (thin, black curves) with experimental data (thick, gray curves)
  corresponding to experiment \#3.  See the caption of Figure
  \ref{KdV2} for axes' definitions.}
\label{KdV3}
\end{figure}

Figure \ref{KdV3} contains plots comparing the KdV predictions with
the experimental data for experiment \#3.  This experiment had an
initial wave amplitude three times of that in experiment \#2 and
therefore it is expected to demonstrate more nonlinear effects than
experiment \#2.  The KdV predictions for this experiment are
qualitatively similar to those in experiment \#2 except that
the amplitudes of the trailing waves are greatly overpredicted.  This
overprediction increases as the waves move away from the wave maker.
These effects may be attributable to dissipative experimental effects
that KdV cannot model.  At each gauge, KdV accurately predicts the
phases for the first couple of waves, but the phase error increases as time increases.  Table \ref{ErrorTable} shows that the error in the KdV predictions for this experiment are an order of magnitude higher than the error in experiment \#2.  Just as in experiment 2, including surface tension slightly improves the KdV prediction.

\subsubsection{The Serre Equations}
\label{SerreNumerics}

\begin{figure}
\centering
\includegraphics[width=16cm]{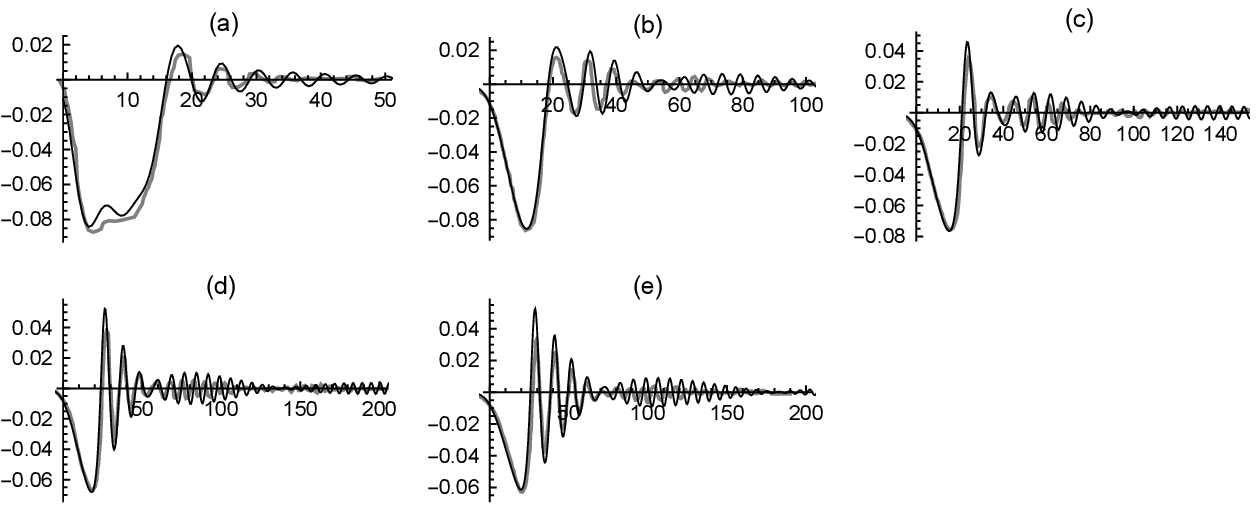}
\caption{Plots comparing results from a numerical solution of the Serre
  equations including surface tension (thin, black curves) with experimental data (thick, gray
  curves) corresponding to experiment \#2.  See the caption of Figure
  \ref{KdV2} for axes' definitions.}
\label{Serre2}
\end{figure}

Figure \ref{Serre2} contains a plot comparing results from numerical
simulations of the Serre equations including surface tension with the data for experiment \#2.
The Serre equations provide a much more accurate prediction for the
experimental data than did KdV.  However, just as with KdV, the (nondisspative) Serre
equations overpredict the amplitudes of the
trailing waves of elevation.  This overprediction increases as the
waves evolve down the tank and may be related to dissipative
effects and the fact that the initial velocity was assumed to be zero.
The Serre equations accurately model the phase of the experimental
data for the initial portions of the time series, but as time
increases at each gauge, the phase error increases.  However, the
phase error is significantly less than in the KdV predictions.
Additionally, the Serre equations predict a group behavior that is
qualitatively similar to what was observed in the experiments.  Comparisons for
experiment \#3 are similar except that the amplitudes of the trailing
waves are significantly overpredicted.  Table \ref{ErrorTable} shows that including surface tension improves the accuracy of the Serre prediction for experiment 3, but not for experiment 2.  Additionally, the Serre equations provide more accurate predictions than does
KdV.  Surprisingly, the ``fully nonlinear'' Serre equations do not
provide a significantly more accurate prediction than KdV for
experiment \#3 even though it is a more nonlinear experiment.  This
may be related to the fact that the unknown experimental initial
velocities were set to zero (see Section \ref{BidWhNumerics} for a
discussion of how velocities impact the model accuracy).

\subsubsection{The Whitham Equation}

\begin{figure}
\centering
\includegraphics[width=16cm]{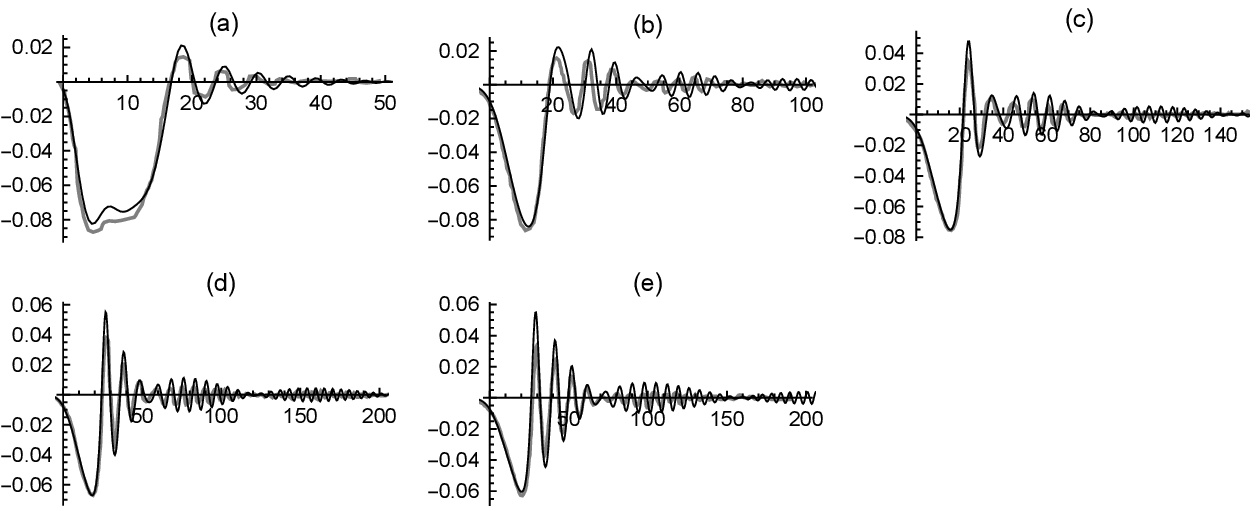}
\caption{A plot of a numerical solution of the Whitham equation
  including surface tension (thin, black curves) and experimental data
  (thick, gray curves) corresponding to experiment \#2.  See the
  caption of Figure \ref{KdV2} for axes' definitions.}
\label{STWhitham2}
\end{figure}

Figure \ref{STWhitham2} demonstrates that the Whitham equation
(including surface tension) does a very good job modeling the wave
evolution in experiment \#2.  It accurately models both the initial
wave of depression and the trailing dispersive waves.  The phases are
accurately reproduced for significant portions of each time series, including
the group behavior of the trailing waves.  This is likely due to the
fact that the phase velocity of the Whitham equation matches the unidirectional phase
velocity of the Euler equations.  Although the (nondissipative) Whitham equation
provides a very good prediction for the initial triangular wave and
phases, it overestimates the amplitudes of most of the trailing
dispersive waves.

\begin{figure}
\centering
\includegraphics[width=16cm]{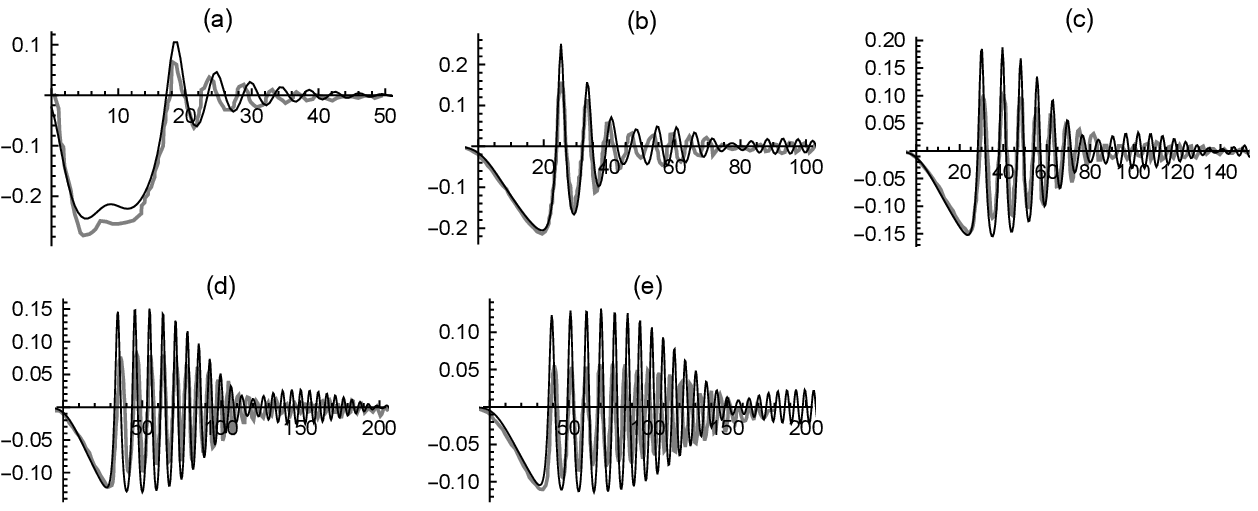}
\caption{A plot of a numerical solution of the Whitham equation
  including surface tension (thin, black curves) and experimental data
  (thick, gray curves) corresponding to experiment \#3.  See the
  caption of Figure \ref{KdV2} for axes' definitions.}
\label{STWhitham3}
\end{figure}

The Whitham results for experiment \#3 are included in Figure
\ref{STWhitham3}.  This figure shows that the Whitham equation models
the initial wave of depression and the phases of many of the trailing waves,
but does not accurately model the amplitudes of the trailing waves
of elevation.  If the amplitude predictions were smaller, then the
error values would be significantly smaller.  Table \ref{ErrorTable}
shows that the Whitham equation provides significantly more accurate
predictions than the KdV and Serre equations for both experiments.
Additionally, the table shows that including surface tension effects
improves the Whitham prediction.

For experiment \#2, the predictions obtained from the Sqrt Whitham
equation are almost identical to those obtained from the Whitham equation.
However there is a significant difference between the Whitham and Sqrt Whitham predictions
for experiment \#3.  This is caused by the fact that experiment \#3 is
more nonlinear than experiment \#2.  The Sqrt Whitham equation
accounts for this more dramatically than does the Whitham equation
because it is a more nonlinear model.  The Sqrt Whitham
equation is less accurate than the Whitham equation for experiment
\#3, see Table \ref{ErrorTable}.  This establishes that the additional nonlinearity in the Sqrt
Whitham equation is not the most appropriate form of nonlinearity
for these experiments.

\subsubsection{The Bidirectional Whitham Equations}
\label{BidWhNumerics}

The bidirectional Whitham models (ASMP, equation (\ref{ASMP}); MKD, equation (\ref{MKD}); HP, equation (\ref{HP})) provide predictions that are similar to one another.  Plots of results from these models are not included due to their similarity to the Whitham results shown in Figures \ref{STWhitham2} and \ref{STWhitham3}.  However, observations can be made by examining the error results contained in Table \ref{ErrorTable}.  First, note that for both experiments the predictions obtained from the bidirectional models all have larger $\mathcal{E}$ values than does the unidirectional Whitham equation.  This is attributed to the fact that the initial horizontal velocities from the experiments were not known and were assumed to be to zero.  The derivation of KdV~\cite{johnson} establishes that the horizontal velocity and surface displacement are related (to the KdV order of accuracy) via
\begin{equation}
u(x,0)=\sqrt{\frac{g}{h_0}}~\eta(x,0),
\label{NonzeroU}
\end{equation}
where $\eta(x,0)$ is the initial surface displacement.  Using (\ref{NonzeroU}) as the initial velocity in the bidirectional Whitham models leads to $\mathcal{E}$ values that are significantly larger than those shown in Table \ref{ErrorTable}.  However, trial and error exploration shows that the $\mathcal{E}$ values for the three bidirectional Whitham models can be decreased to less than $0.06$ for experiment \#2 and less than $0.9$ for experiment \#3 if the velocities are chosen to be proportional to the initial surface displacement (with a constant of proportionality less than $\sqrt{\frac{g}{h_0}}$).  This suggests that the bidirectional Whitham equations would outperform the Whitham equation and would be the most accurate of the models examined if the initial horizontal velocities were known.  To truly answer this question, similar experiments in which the horizontal velocities are measured must be conducted.

\medskip

\noindent{\textbf{Additional observations:}}
\begin{itemize}
\item{All bidirectional Whitham systems very accurately model the
    experimental phases because they exactly reproduce the Euler
    phase velocity.}
\item{All bidirectional Whitham systems accurately model the initial
    wave of depression.}
\item{Including surface tension typically improves the accuracy of the
    predictions, regardless of the model.}
\item{The ASMP and HP systems return very similar predictions.  This is an interesting result given that the ASMP system is suspected to be ill-posed if the surface displacement is negative~\cite{EhrnstromPeiWang,Kyle} as it is here.  Our simulations showed no sign of singularity even though we computed over intervals longer than the experimental data required.  This result may be related to the fact that Claassen \& Johnson's numerics show that the more negative the initial surface displacement, the more rapidly the singularities arise.  Only a small percentage of the initial surface displacement in equation (\ref{ICs}) is negative and therefore it is possible that the singuarity, if one exists, arises outside the interval we examined.}
\item{As with all of the other models, the bidirectional Whitham
    equations do not accurately model the amplitudes of the trailing
    waves.  The amplitude overprediction may be related to the facts
    that all of the models considered herein are nondissipative and the experimental data
    appears to demonstrate dissipative effects.  Many models of
    dissipation for small-amplitude, long waves have been developed,
    including the KdV-Burgers system (e.g.~\cite{Whithambook,El}) and dissipative Boussinesq-type systems (e.g.~\cite{Keul,Mei95,DutykhDias2007,ChenGoubet2007}).
    Although this is an interesting, open question, these models and
    effects are outside the scope of this work because the goal here is to compare the accuracy of the recently derived (conservative) bidirectional Whitham models.}
\item{The higher-order MKD system provided predictions that are more
    accurate than the other bidirectional models and almost as
    accurate as the Whitham equation even when the initial horizontal
    velocity is assumed to be zero.  By choosing the initial
    velocity to be proportional to the surface displacement, the MKD
    system provides the most accurate approximation to the
    experimental data.  This suggests that nonlinearity of MKD form is
    necessary and that the leading-order nonlinearity utilized by the
    KdV and Whitham equations is not sufficient to accurately model
    these experiments.}
\end{itemize}

\section{Summary}
\label{summary}

We have compared laboratory measurements from two surface water-wave experiments on shallow water with predictions from the KdV, Serre, Whitham, and several bidirectional Whitham equations.  We showed that predictions obtained from the Whitham equation (with or without surface tension) were significantly more accurate than those obtained from the KdV and Serre equations (with or without surface tension) for these experiments.  The Whitham and bidirectional Whitham equations very accurately modeled the phases observed in the experiments.  This is attributable to the fact that they all have accurate reproductions of the Euler phase velocity.  We showed that nonlinearity played an important role in these experiments.  The predictions obtained from the linear theory provided the least accurate approximations to the experimental data.  We showed that including surface tension typically improved the accuracy of the predictions.  Finally, we found that the bidirectional Whitham models provided predictions that were almost as accurate as the Whitham equation and we hypothesize that, had the initial horizontal velocities been known, the bidrectional Whitham predictions would have been the best.  The success of the bidirectional Whitham equations in modeling these experiments suggests that they are appropriate to model real-world phenomena.

\section*{Acknowledgments}

The author thanks Vincent Duchene, David George, Philippe Guyenne, Diane Henderson, and Harvey Segur for helpful discussions and comments.  Special thanks to Knox Hammack for digitizing the data from~\cite{HamSeg}.  The author was supported by the National Science Foundation under grant DMS-1716120.  This material is based upon work supported by the National Science Foundation under Grant No.~DMS-1439786 while the author was in
residence at the Institute for Computational and Experimental Research
in Mathematics in Providence, RI, during the Spring 2017 semester.

\end{document}